\documentstyle[multicol,aps,prl,epsf,psfig]{revtex}
\begin{document}
\def\Dscr{{\cal D}}
\def\DsC{{\cal C}}
\def\DsS{{\cal S}}
\def\DsN{{\cal N}}
\def\DsE{{\cal E}}
\renewcommand{\textfraction}{0.0}
\renewcommand{\topfraction}{1}
\renewcommand{\bottomfraction}{1}
\setcounter{topnumber}{50}
\setcounter{bottomnumber}{50}
\setcounter{totalnumber}{50}
\setlength{\floatsep}{\baselineskip}
\setlength{\textfloatsep}{\baselineskip}
\renewcommand{\thefigure}{\arabic{figure}}
\topmargin = +.1 in 

\title{Towards A Consistent Modeling Of Protein
Thermodynamic And Kinetic Cooperativity:
How Applicable Is The Transition State Picture To
Folding and Unfolding?}

\author{H\"useyin KAYA  and Hue Sun CHAN}
\address{Department of Biochemistry, and \\
Department of Medical Genetics \& Microbiology \\
Faculty of Medicine, University of Toronto \\
Toronto, Ontario M5S 1A8, Canada}

\maketitle

\begin{abstract}

To what extent do general features of folding/unfolding kinetics of small 
globular proteins follow from their thermodynamic properties? To address
this question, we investigate a new simplifed protein chain model
that embodies a cooperative interplay between local conformational 
preferences and hydrophobic burial. The present four-helix-bundle 55mer 
model exhibits proteinlike calorimetric two-state cooperativity. It 
rationalizes native-state hydrogen exchange observations. Our
analysis indicates that a coherent, self-consistent physical account of 
both the thermodynamic and kinetic properties of the model leads naturally
to the concept of a native state ensemble that encompasses considerable 
confomational fluctuations. Such a multiple-conformation native state 
is seen to involve conformational states similar to those revealed by 
native-state hydrogen exchange. Many of these conformational 
states are predicted to lie below native baselines commonly used in 
interpreting calorimetric data.  Folding and unfolding kinetics are 
studied under a range of intrachain interaction strengths as in 
experimental chevron plots. Kinetically determined transition midpoints 
match well with their thermodynamic counterparts.
Kinetic relaxations are found to be essentially single exponential 
over an extended range of model interaction strengths. This includes the 
entire unfolding regime and a significant part of a folding regime 
with a chevron rollover, as has been observed for real proteins that
fold with non-two-state kinetics. The transition state picture of
protein folding and unfolding is evaluated by comparing 
thermodynamic free energy profiles with actual kinetic rates.
These analyses suggest that some chevron rollovers may arise from an 
internal frictional effect that increasingly impedes chain motions with 
more native conditions, rather than being caused by discrete deadtime folding 
intermediates or shifts of the transition state peak as previously posited.
\\

\noindent
{\bf Running title:} Transition State Picture of Protein Folding


\end{abstract}

\begin{multicols}{2}

\section{INTRODUCTION}

Our physical knowledge of protein folding is measured by the 
extent to which current ideas of elemental polypeptide interactions 
are capable of reproducing experimental data. 
Tremendous experimental progress has been made in the
recent past.$^{1-6}$ During the same time, 
investigations of simplified self-contained polymer models 
have contributed much physical insight.$^{7-16}$
These models are successful in physically rationalizing many general 
features of proteins in terms of polymer
properties, building a foundation for future advances. To 
move forward in theoretical development, it is necessary to 
recognize what common protein properties have not been predicted by models 
to date and target our efforts towards rectifying such deficiencies.
A prime example is the current lack of chain models that can quantitatively 
reflect the extreme kinetic and thermodynamic cooperativity of small 
single-domain proteins.$^{5,17}$ This highlights 
our insufficient understanding of protein energetics even at a 
``big-picture'' level, suggesting that as heteropolymers natural proteins 
may be quite special in some respects.

We have recently investigated the severe constraints imposed on protein polymer
models by the experimental observations of calorimetric and other hallmarks
of thermodynamic two-state cooperativity.$^{18-20}$ 
These cooperativity requirements appear to be more stringent than
most other generic protein properties studied so far. A variety of flexible 
heteropolymer models with additive residue-based contact energies 
are able to explain significant aspects of the folding process.$^{7-16}$
But such additive models --- at least for the
several examples evaluated to date --- are found to be insufficient to 
satisfy the thermodynamic cooperativity requirements, even though deviations 
from proteinlike thermodynamics can be lessened in some instances by 
enhancing interaction heterogeneity through using larger numbers of letters 
and repulsive energies in model alphabets.$^{19}$ As far
as thermodynamic cooperativity is concerned, in scenarios 
tested thus far, we find that proteinlike thermodynamics can arise from 
nonadditive cooperative contributions that originate from an interplay 
between local conformational preferences and (mostly nonlocal) interactions 
that favor formation of protein cores.$^{18,20}$

In the folding literature, an intimate correspondence between protein 
thermodynamics and kinetics has figured prominently in 
theoretical,$^{7,21-23}$ modeling,$^{24-27}$
and interpretative$^{2,4}$ discourses. Therefore, we ask:
Given that a heteropolymer model has already been constrained to satisfy
a set of proteinlike thermodynamic properties, to what extent proteinlike 
kinetic behavior would follow automatically? For instance, would
such a model be sufficient for two-state kinetics as observed for 
many small single-domain proteins$^{5}$? More generally, 
what improvement in proteinlike kinetics would such a model enjoy over other 
models that are now known to be thermodynamically less cooperative? 

In analyses of folding/unfolding kinetics experiments, free energy profiles 
are used extensively to provide useful insight$^{2,4,28-33}$
and as a picturesque device to summarize data.  However, 
other than the folding free energy and rate measurements themselves, 
independent experimental techniques to accurately define and determine such 
profiles are currently lacking. Moreover, many of these profiles were
proposed without considering explicit chain representations. Therefore, the 
applicability and generality of their implied physical pictures remain to be 
ascertained. Coarse-grained protein chain models are well suited to shed light 
on this fundamental issue because they allow for broad conformational sampling.
Free energy profiles in coarse-grained models can be 
obtained directly from chain population distributions, without regard to 
(and therefore independent of) kinetic rates. It follows that a rigorous 
evaluation of the applicability of transition state theory to protein 
folding can be conducted by comparing the transition-state-predicted 
rates and the actual kinetic rates in these models. We study one such model 
below.

\section{\bf A MODEL FOR THERMODYNAMIC COOPERATIVITY}

\noindent
The present analysis is based on a thermodynamically cooperative 
55mer lattice protein model that folds to a ground state with
a four-helix core (Fig.~1). The intrachain interaction
scheme includes additive 5-letter contact energies,$^{20,34}$ 
repulsive interactions disfavoring left-handed helices and sharp turns at the 
end of a helix, as well as cooperative ``native hydrogen bond burial'' 
terms$^{20}$ (c.f. refs.~35, 36). 
The total energy $E$ is defined by Eq.~(1) in ref.~20.
Although ``native-centric'' interactions were introduced to enhance 
thermodynamic cooperativity in the present model, unlike the usual G\=o 
construction, they are not necessary for recognizing the ground state.
This is because the general, non-native-centric terms in the model 
(all terms in $E$ except the ``native hydrogen bond burial'' terms) are
sufficient to provide global favorability to the proteinlike 
four-helix ground state. We note that several other studies$^{36-39}$ 
have also emphasized cooperative interactions in protein folding; and
nonadditive aspects of hydrophobic effects are being explored.$^{40-45}$ 
As we have emphasized,$^{20}$
although the present model is useful for exploring the issues at hand,
it should be regarded as tentative, partly because it does not provide an
explicit account of other possible physical origins of protein 
thermodynamic cooperativity such as sidechain packing.$^{19,46,47}$
Furthermore, in view of the current lack of definitive understanding of 
hydrogen bonding energetics (see discussion and references in ref.~20), 
the cooperative ``native hydrogen bond burial'' energy in the present model 
should be broadly interpreted as representative of a general favorable 
coupling between local conformational preference and formation of proximate 
tertiary contacts, the physical mechanisms of which remain to be
further elucidated.

\begin{figure}
\begin{center}
\leavevmode
\psfig{figure=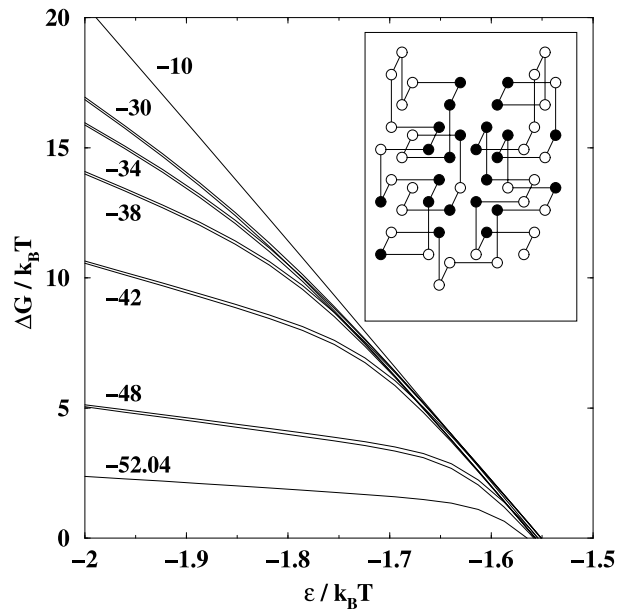,width=8cm,height=8cm,angle=0}
\end{center}
\narrowtext   
\caption{Thermodynamic stabilities and definitions of native
and denatured states. One of the eight iso-energetic ground-state
conformations is shown in the inset, where black beads denote
nominally hydrophobic residues.$^{20}$ Free energies of
unfolding $\Delta G=k_B T\ln(P_{\rm N}/P_{\rm D})$.
Solid curves (labeled by $E_t$) classify conformations with
$E\le E_t$ and $E>E_t$ as native and denatured, respectively.
Dashed curves show the free energy of denatured conformations,
defined as those with $E$ greater than the values shown, relative
to the ground-state-only native state with $E_t=-52.04$. All results
presented in this paper were obtained using model energetic parameters
$\gamma_{\rm lh}=6.0$, $\gamma_{\rm st}=5.0$, ${\cal E}_{\rm Helix}=0$,
$b{\cal E}_{\rm Hb}=-0.8$, and $b=1.5$ as specified ref.~20.}
\label{step}
\end{figure}

In addition to the formulation in ref.~20,
here we introduce a parameter $\epsilon$ to model protein behaviors at 
different intrachain interaction strengths, such that the {\it effective}
energy of a conformation with energy $E$ is equal to $-\epsilon E$, hence
its Boltzmann weight equals $\exp(\epsilon E/k_B T)$, where $k_BT$ is
Boltzmann's constant times absolute temperature. It follows that
the partition function $Q=\sum_{E}g(E)\exp(\epsilon E/k_B T)$, 
where $g(E)$ is the number of conformations with 
energy $E$, and $g(E)$ is estimated by a parameter-space Monte Carlo 
histogram technique.$^{20}$
The formulation in ref.~20 corresponds 
to $\epsilon=-1$. Because of the peculiar and significant
temperature dependence of the solvent-mediated interactions in real proteins,
varying $-\epsilon/k_B T$ serves better as a model for how
effective intrachain interactions are modulated at constant
temperature by denaturant concentration$^{48}$ or denaturant activity$^{37}$ 
than for how Boltzmann weights changes with temperature.$^{11,49-51}$
Here, as a first approximation, denaturant effects are simply taken to be 
uniform over different interaction types. Models with different 
denaturant effects on different interaction types remain to be explored.

\begin{figure}
\begin{center}
\leavevmode
\psfig{figure=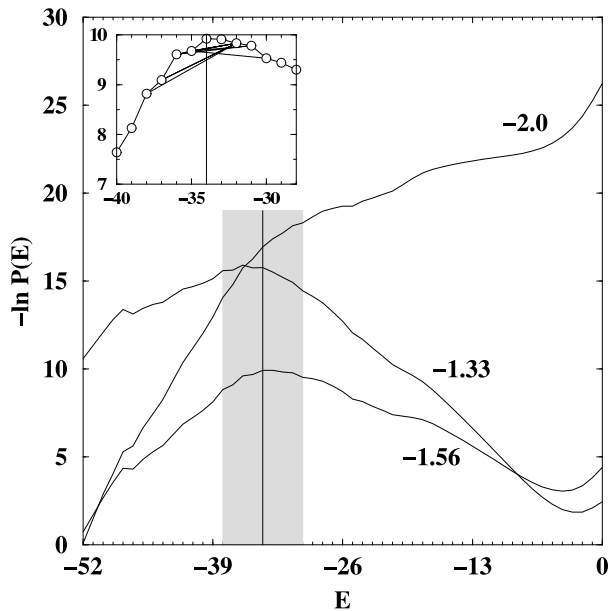,width=8cm,height=8cm,angle=0}
\end{center}
\narrowtext
\caption{Free energies profiles are given by negative logarithmic distributions 
of energy (solid curves), plotted here for the $\epsilon/k_B T$'s shown.
$P(E)$ is the sum of Boltzmann weights of conformations
with energies $E^\prime$ in the range $E-0.5<E^\prime\le E+0.5$.
The vertical dashed line marks the $E=-34$ free energy peak for
$\epsilon/k_B T=-1.56$. The inset shows the peak region of this profile,
where lines joining a pair of open circles [$-\ln P(E)$ values] record
all single-chain-move interconversions between a conformation with $E<-34$
and one with $E>-33$ in our simulation. These kinetic
connections suggest identifying the shaded area ($-38\le E \le -30$) 
as a transition state region. }
\end{figure}

As we have discussed from a general polymer perspective,$^{19}$
matching theoretical considerations with the experimental practice of 
calorimetric baseline subtractions
necessitate a multiple-conformation native state that entails considerable 
fluctuations beyond small-amplitude 
vibrations. Here we further investigate the implications of native-state 
conformational diversity. To that end, we study different definition
of ``native'' and ``denatured'' states by assigning different values for a 
``transition'' energy $E_t$ demarcating the native and denatured ensembles,
such that $P_{\rm N}$ $=\sum_{E\le E_t}g(E)\exp(\epsilon E/k_B T)/Q$
is the fractional native population and $P_{\rm D}=1-P_{\rm N}$ 
is the fractional denatured population (Fig.~1). 
For each of these definitions to be tested, a range of energies is spanned by 
the native state, except for the special case when $E_t$ is chosen to be 
equal to the ground-state energy.
Remarkably, despite the differences in the definitions of ``native'' 
and ``denatured'' states, 
the thermodynamic ($\Delta G=0$) transition midpoints of the different $E_t$'s 
in Fig.~1 are very similar, all at $\epsilon/k_B T\approx -1.56$. 

Stabilities of different denatured ensembles relative to that of the 
ground state are shown in Fig.~1 (dashed curves). These quantities
correspond roughly to native state hydrogen 
exchange (HX) free energies,$^{52-54}$ (see also ref.~55),  for it
is reasonable to expect that certain amides become exposed and 
exchangeable when the effective energy of a conformation is 
above a certain threshold. Our results share the same general trend as 
that observed in these experiments,$^{52-54}$
suggesting that some of the fluctuations 
observed by HX may be considered as part of a multiple-conformation 
native state.$^{18,19,56}$
Figure~1 indicates that 
linear extrapolation of $\Delta G$ from the transition region to the strongly 
native regime (more negative $\epsilon/k_B T$) is only valid for the 
free energy difference between the set of ``fully unfolded'' open conformations 
and the ground state (top dashed curve).

\section{\bf FREE ENERGY PROFILES AND CHEVRON PLOTS}

\noindent
Consistent with the model's thermodynamically two-state character,$^{20}$
its energy distribution is bimodal under denaturing conditions and moderately 
native conditions (Fig.~2), although the 
distribution becomes one-sided or ``downhill''$^{7,57,58}$ 
under strongly native conditions (e.g., 
when $\epsilon/k_B T=-2.0$). We emphasize that here the determination of 
$\ln P(E)$ is entirely independent of any kinetic consideration. Therefore, 
the present $\ln P(E)$ function reflects the actual thermodynamics of the 
model. As such, its physical origin is fundamentally different from 
free energy profiles that have been empirically constructed or postulated 
to fit rate data. We can therefore use these $\ln P(E)$'s to 
assess the transition state picture, with $E$ as the reaction 
coordinate. Different reaction coordinates have been used in other 
investigations.$^{59,60}$

We employ standard Metropolis Monte Carlo dynamics$^{19,22}$ to explore 
physically plausible kinetic scenarios, using the number of attempted moves 
as the model time. This approach has been proven useful$^{7-11,13,61}$
despite its obvious limitations.$^{11,16,50}$
The present set of elementary chain moves 
consists of end flips, corner flips, crankshafts,$^{59-61}$ 
rigid rotations,$^{11}$ 
and local moves that transform two turns of a right-handed helix among 
its three possible orientations while holding its two end monomers fixed. 
The relative frequencies of attempting 
these moves are 2.3\%, 27\%, 60.6\%, 10\%, and 0.1\%, respectively.
Some chain moves can lead to large changes in energy, hence movements along 
the model free energy profile need not be continuous (inset of Fig.~2).
Therefore, it is more justified to regard the transition state as a region 
rather than a single highest point along this particular
reaction coordinate.$^{22,26,60}$ The group of conformations represented
by the shaded area in Fig.~2 clearly serves the role of a transition state 
because all conformational interconvertions between the native and 
unfolded sides of the population distribution must pass through 
one or more conformations in the shaded area.

\begin{figure}
\begin{center}
\leavevmode
\psfig{figure=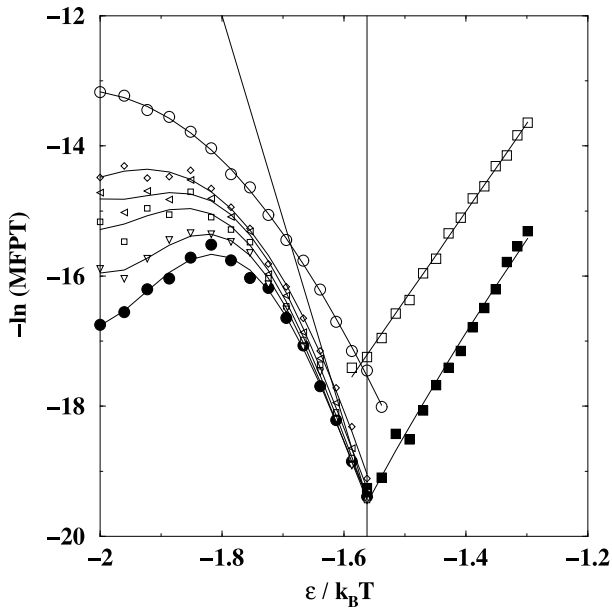,width=8cm,height=8cm,angle=0}   
\end{center}
\narrowtext
\caption{Model chevron plots. Average logarithmic rates are given by negative
logarithms of mean first passage time (MFPT). Each folding trajectory
starts from a randomly generated conformation; unfolding trajectories are
initiated from the ground state.  Each data point is averaged from
$\sim 50$--$1,000$ trajectories. Solid curves through data points
are mere guides for the eye. Larger squares on the right show unfolding
MFPT's for attaining energies $E>-34$ (open) and $E>-4$ (filled).
Unfolding MFPT's for $E>-10$ are essentially identical to that for $E>-4$.
Other data points (on the left) show folding MFPT's for reaching
({\it from top to bottom}) $E\le -34$, $-40$, $-42$, $-44$, $-46$, and
the ground state. The vertical dashed line is the approximate
transition midpoint. The inclined dashed-dotted line shows folding rates
if kinetics were two-state for the present model, an hypothetical
situation in which the ground state thermodynamic stability relative to the
fully unfolded conformations (c.f. dashed line labeled by ``$-10$''
in Fig.~1) is given by the difference between logarithmic folding rates 
(dashed-dotted line) and unfolding rates extrapolated from the solid squares. }
\end{figure}

Figure~3 reports simulated mean first passage times$^{59,60}$ for
a range of intraprotein interaction strength on both sides of the 
transition midpoint, in a format identical to typical 
experimental chevron plots.$^{29,31,32,49,51,62}$
We explore a variety of kinetic folding and unfolding criteria by 
monitoring the time it takes for the chain to first cross several 
different ``finish lines.'' This results in an appreciable variation in 
apparent rates under strongly native conditions (Fig.~3, more negative 
$\epsilon/k_B T$). Similar effects may be operative when multiple 
experimental probes are used to monitor kinetics.$^{63-65}$

The trajectory in Fig.~4 (upper panel) depicts the model's 
heuristically ``two-state'' behavior at the transition midpoint. For the 
two chain properties shown, native and denatured parts of the trajectory 
can be easily discerned, with very little time spent in between; strongly 
suggesting that the kinetics is first order. Fluctuations in $E$ is 
considerable within the native (low $E$) part of the trajectory, underscoring
the utility and necessity of a multiple-conformation native state (see 
Fig.~1 and below). Another facet of the native-denatured interconversion 
is provided by the $R_g$ trace. Consistent with a recent kinetic $R_g$ 
measurement on a small protein,$^{66}$
it shows that the chain undergoes 
sharp kinetic transitions between a native state that has minimal fluctuations 
in $R_g$ and a denatured state that spans a wide range of $R_g$'s.

A more quantitative test introduced by Gutin et al.$^{67}$ is performed in 
the lower panel of Fig.~4. It indicates that folding kinetics is essentially 
first order (i.e., single-exponential) for an extended range of 
model intraprotein interaction strength, covering moderately native conditions 
($\epsilon/k_B T\approx -1.80$) through conditions that are less favorable 
to folding (less negative $\epsilon/k_B T$), although deviations from 
single-exponential behavior occur under strongly native conditions 
in the model ($\epsilon/k_B T< -1.85$). 
Using the same technique, unfolding kinetics 
(Fig.~3) is found to be essentially single-exponential for the entire 
range of unfolding $\epsilon/k_B T$ investigated (detailed results not shown). 
We have confirmed these conclusions by analyzing first passage time (FPT) 
distributions as in ref.~68 at
several $\epsilon/k_B T$'s, paying special attention to folding kinetics under 
moderately native conditions that are not far from the onset of drastic
chevron rollover and non-single-exponential behavior (Fig.~3). For example, 
we have obtained the logarithmic FPT distribution at $\epsilon/k_B T=-1.72$ 
by binning 1,080 simulated trajectories into time slots of $10^6$, and
found that 98\% of these trajectories can be fitted by a single exponential 
with a correlation coefficient $r=0.95$. 
If one assumes that the unit model time needed for each elementary chain 
move corresponds roughly to a real time scale
of $10^{-11}$--$10^{-9}$ sec (ref.~69),
the fastest model folding rate in Fig.~3 is in the 
order of $10^2$--$10^4$ sec$^{-1}$.

The contrast between the present model and its corresponding G\=o model is
intriguing. We have shown that the G\=o model in Fig.~4 is 
thermodynamically significantly less cooperative,$^{20}$ yet 
it folds faster than our model. This scenario of 
a {\it negative} correlation between folding speed and thermodynamic 
cooperativity may bear on the issue of folding rate overestimation 
in folding theories that use G\=o-like potentials.$^{23}$
It also raises a more basic question as to whether and when
the G\=o prescription is sufficiently adequate for capturing minimal 
frustration$^{21}$ mechanisms in real proteins.

\begin{figure}
\begin{center}
\leavevmode
\psfig{figure=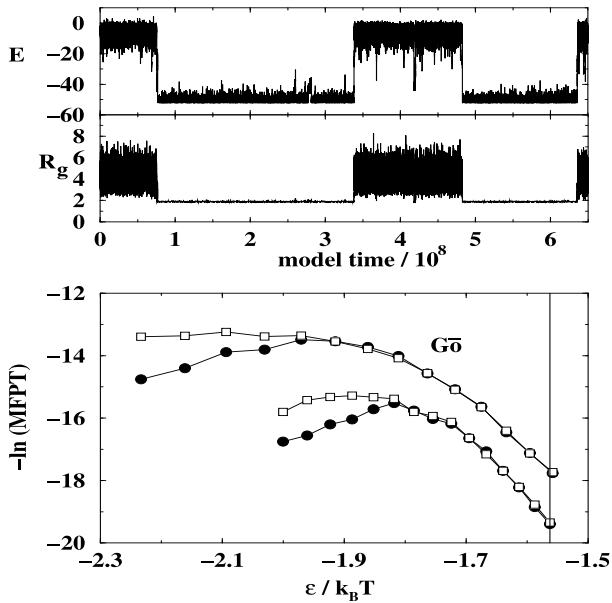,width=8cm,height=8cm,angle=0}
\end{center}
\narrowtext
\caption{{\bf Upper panel:} A typical trajectory at
$\epsilon/k_B T=-1.56$. $R_g$ is radius
of gyration in units of lattice bond length.
{\bf Lower panel:} Folding MFPT's (filled circles) through the ground
state for the model in Fig.~3 (lower curves) are compared to that for
a G\=o model (upper curves) that has the same transition midpoint
(vertical dashed line), uses the same move set, and assigns a $-1.5$
energy for every contact in the conformation in Fig.~1 and zero
energy otherwise. Lines through
data points are mere guides for the eye. Open squares are median first
passage times divided by $\ln 2$, which equals MFPT for
single-exponential kinetics. Hence a discrepancy between the circles and
squares signals a deviation from single-exponential kinetics.$^{67}$}
\end{figure}

The present model is proteinlike in that it predicts a mild chevron rollover 
concomitant with single-exponential folding kinetics, consistent 
with experiments.$^{29,31,62}$ But the drastic chevron 
rollover (with an appreciable decreasing folding rate with increasing 
native conditions, and non-single-exponential folding kinetics)
predicted for strongly native conditions in the present model 
($\epsilon/k_B T< -1.85$ in Fig.~3) has not been documented for real 
proteins. This suggests that such conditions, which coincide with
downhill folding$^{7,57,58}$ in the present model (see above), may not
be realizable. If so, this is not surprising. Native 
stability can be arbitrary high in the model ($-\epsilon/k_B T$ can be 
arbitrarily large), but for real proteins native stability is limited 
by the actual chemistry at zero denaturant. It follows that the experimental 
zero-denaturant situation for most proteins most likely corresponds 
to $\epsilon/k_B T> -1.80$ in Fig.~3.
It would be interesting to explore whether special experimental situations 
corresponding to the very strongly folding conditions in the model 
can be found for some proteins such that similar drastic chevron rollovers 
can be observed.

\section{\bf ASSESSING THE TRANSITION STATE PICTURE}

\noindent
Consistent with the thermodynamics of our model, kinetic rate of folding 
to the ground state and of unfolding to an open state meet at the 
approximate transition midpoint determined thermodynamically
in Fig.~1 (c.f. the lower ``V'' in Fig.~3). 
Interestingly, a similar consistency is seen by matching
rates of crossing the peak of the free energy profile 
in the inset of Fig.~2 in the folding and unfolding
directions (upper ``V'' in Fig.~3). Near the transition midpoint, 
folding rates defined by crossing several different finish lines at low 
but non-ground-state energies are very close to one 
another, and are only slightly faster than the rate of folding to the ground 
state (Fig.~3). This implies that, under midpoint to moderately native 
conditions, kinetics is rapid once the folding chain has cleared the 
shaded transition state region in Fig.~2 and proceeds on to the native side. 
But the folding rates for different finish lines
are very different under strongly native conditions ($\epsilon/k_B T\approx
-2.0$), indicative of glassy dynamics (Figs.~3 and 4).

Despite the essentially single-exponential and heuristically ``two-state'' 
kinetics discussed above, the folding/unfolding kinetics of the present model 
differs from the strictly two-state variety observed for an increasing 
number of small single-domain proteins$^{4,5,24,25,70}$
 such as a 64-residue form of 
chymotrypsin inhibitor 2.
Figure~3 shows that folding rates under moderately to 
strongly native conditions are slower than that required for such strictly 
thermodynamically {\it and} kinetically two-state proteins (inclined 
dashed-dotted line). In fact, Fig.~3 is reminiscent of experimental 
chevron plots with rollovers.$^{31,32,71}$ Examples include wildtype 
barnase$^{29}$ and ribonuclease A.$^{62}$
Hence we believe 
the present lattice construct may serve as a tool for better understanding 
the folding kinetics of these proteins.

How much kinetic information can be inferred from free energy profiles 
such as those in Fig.~2? The conventional transition state picture of
protein folding$^{2,28,30}$ stipulates that
$$
{\rm rate} = F \exp\biggl(-{\frac {\Delta G^\ddagger} {k_B T}}\biggr) \; ,
\eqno(1)
$$
where $\Delta G^\ddagger$ is the activation free energy for the kinetic
process in question. We call $\Delta G^\ddagger/k_B T$ the 
transition-state exponent. $F$ is the pre-exponential front factor$^{11}$
or prefactor,$^{23}$ which depends on solvent viscosity (not considered
here) but is often taken to be insensitive to intraprotein interaction 
strength.$^{30}$ Figure~5 examines whether this picture 
applies to the present model. It does so by investigating the dependence of $F$ 
on $\epsilon/k_B T$. For the sake of generality, several physically 
motivated $P(E)$-based definitions of ``transition state,'' ``folded state'' 
and ``unfolded state'' are evaluated. In the tests conducted here,
``transition state'' is defined by either the shaded area in Fig.~2 
($-38\le E\le -30$) or $E=-34$ (peak of barrier); 
``folded state'' is defined by the ground state only ($E=-52$) or by 
$E\le -34$ (left of the barrier); and ``unfolded state'' is defined by 
the approximate position of the denatured free energy minimum ($E=-4$), 
by the bulk of the open conformations ($E>-10$), or by $E\ge -34$ 
(right of the barrier).
Transition-state exponents for folding and unfolding are then computed,
respectively, by the [transition/unfolded] and [transition/folded]
population ratios at the given $\epsilon/k_B T$ (Fig.~5).

The scaling of $F$ with respect to $\epsilon/k_B T$ is found to be
not sensitive to these variations in definition (Fig.~5, upper panel). 
Our results show that the simple transition state picture does not apply 
to folding in this model. For the quasi-linear part of the chevron 
plot in Fig.~3, the relationship between $\Delta G^\ddagger/k_B T$ 
and $-\ln({\rm MFPT})$ is approximately linear (Fig.~5), with slope
$\approx -1.5$ (filled symbols) or $\approx -1.9$ (open symbols).
This implies that $F$ has approximately the same functional form as the 
activation factor in this regime, but with an exponent of opposite sign, viz.,
$F\sim \exp(-\mu\Delta G^\ddagger/k_B T)$, where $\mu$ $\approx -0.33$
for folding to the ground state (filled symbols). On the other hand, 
unfolding appears to be well described by the simple transition state 
picture. The corresponding slope for unfolding $\approx -1.0$, hence 
$F\approx$ constant. Folding and unfolding front factors are approximately 
equal near the thermodynamic midpoint (Fig.~5, lower panel), reflecting
the fact that in that region folding and unfolding rates are essentially 
equal (Fig.~3). The role of front factors in understanding 
folding rates has recently been emphasized by Portman et al.$^{23}$
The pattern in Fig.~5 is similar in some respects to the results of
a recent model study by Nymeyer et al.$^{22}$ Using a different reaction 
coordinate for 2-, 3-letter and G\=o 27mer models, these authors found 
approximate linear relations with non-unity slopes between kinetically 
simulated rates and rate quantities deduced from free energy profiles, although 
they did not consider a broad range of overall interaction strengths 
as that in chevron plots.

Figure~5 shows that $F$ for folding decreases with increasingly native 
conditions. $F$ may be identified as an effective diffusion coefficient.
It corresponds to an internal friction term arising from the 
impediment to motion imposed by the chain segments on one 
another.$^{2,7,23,60,71-74}$ 
Plaxco and Baker$^{72}$ have experimentally investigated 
internal friction in protein folding, and concluded insightfully that internal
friction effects are limited for strictly two-state proteins. But the 
functional form they considered is different from the novel approximate 
exponential form $F\sim \exp(-\mu\Delta G^\ddagger/k_B T)$ discovered here 
for the quasi-linear part of chevron plots and the part with a mild
rollover (corresponding to a limited range 
of $\epsilon/k_B T$) for proteins that are not kinetically two-state. 
In the present model, when a wider range of $\epsilon/k_B T$ is considered 
(Fig.~5, lower panel), the folding $\ln F$ reveals 
its nonlinear character, a feature anticipated by energy landscape 
theory$^{21,23}$ and consistent with a pioneering simulation 
study of Socci et al.$^{60}$

\begin{figure}
\begin{center}
\leavevmode
\psfig{figure=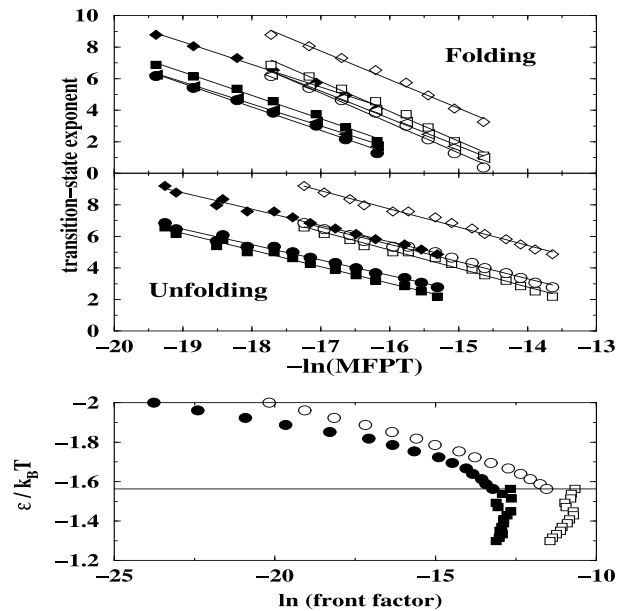,width=8cm,height=8cm,angle=0}
\end{center}
\narrowtext
\caption{Correlations between rates and transition-state
exponents. Filled symbols are for reaching the ground state in folding and
reaching $E>-4$ in unfolding, open symbols are for crossing the
$-\ln P(E)$ peak at $E=-34$ in either directions.
{\bf Upper panel:} Folding transition-state exponent
$\Delta G^\ddagger/k_B T=$ $-\ln[P(-38\le E\le -30)/P(E\ge -10)]$ (circles),
$-\ln[P(-38\le E\le -30)/P(E\ge -34)]$
(triangles), $-\ln[P(E=-34)/P(E=-4)]$ (squares), or
$-\ln[P(E=-34)/P(E\ge -10)]$
(diamonds). Data points shown are for $\epsilon/k_B T\ge-1.75$.
For unfolding, $\Delta G^\ddagger/k_B T=$
$-\ln[P(-38\le E\le -30)/P(E=-52)]$ (squares),
$-\ln[P(-38\le E\le -30)/P(E\le -34)]$
(circles), or $-\ln[P(E=-34)/P(E=-52)]$ (diamonds).
The straight lines are fitted.
{\bf Lower panel:} $\ln F$ $\equiv$ $-\ln({\rm MFPT})$ $+$
$\Delta G^\ddagger/k_B T$ (horizontal variable) vs.
$\epsilon/k_B T$. Filled and open
circles (for folding) and squares (for unfolding) identify
the folding and unfolding $\Delta G^\ddagger/k_B T$ used, as defined in the 
upper panel. The dashed line marks the approximate transition midpoint.}
\end{figure}

\section{\bf THE CALORIMETRIC CONNECTION: WHAT IS THE NATIVE STATE?}

\noindent
A thermodynamic consideration of the model's free energy 
profiles (Fig.~2) and the above kinetic analysis suggest that 
a natural way to define the ``native'' and ``denatured'' states
is to have their demarcation line at $E_t=-34$. Figure~6 investigates
the calorimetric implications of different $E_t$ choices. As a first 
test of principles, in this section we take the effective intraprotein 
energies as temperature independent. Our conclusions are not expected 
to be significantly altered by the incorporation of proteinlike
temperature-dependent effective interactions.$^{18}$

\begin{figure}
\begin{center}
\leavevmode
\psfig{figure=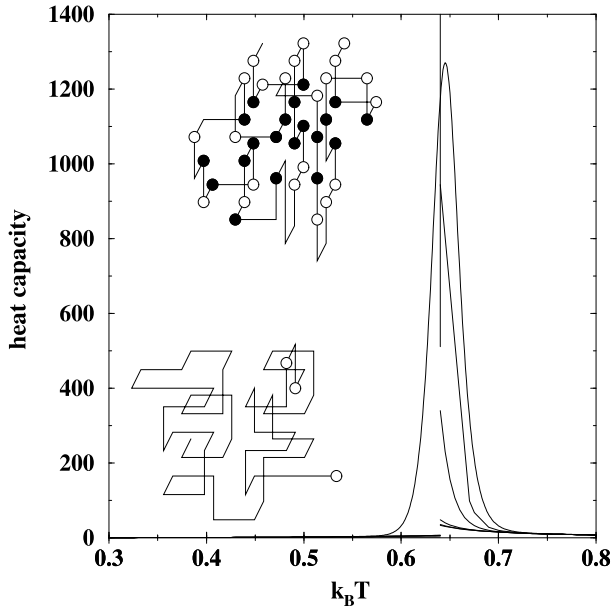,width=8cm,height=8cm,angle=0}
\end{center}
\narrowtext
\caption{Heat capacity vs. temperature (from ref.~20).
Here nonlinear denatured (high $T$) dashed baselines$^{19,75}$ are computed
for ({\it from top to bottom}) $E_t=$ $-52.04$, $-48$, $-42$, $-38$,
and $-34$ (c.f. Fig.~1). Corresponding native baselines are plotted but
are too close to one another to be distinguishable. The vertical
dotted line marks the approximate transition midpoint used throughout
this work. Also shown are example compact non-ground-state conformations
with $E=-36.0$ (top) and $E=-46.0$ (bottom). The beads mark monomers
that are not in their folded environment, i.e., do not have their full
set of contacts as that in the ground state.}
\end{figure}

We have argued that empirical calorimetric baseline subtractions correspond 
essentially to an operational definition of the native and denatured 
ensembles.$^{19}$
The demarcation energy (or enthalpy) between the native 
and denatured states may be ascertained by matching empirical baselines$^{19}$
to the nonlinear ``formal two-state'' baselines of Zhou et al.$^{75}$
Figure~6 shows that the baselines from empirically extrapolating the native 
and denatured tails of the heat capacity curve of the present model$^{20}$
essentially coincide with the formal two-state baselines for $E_t=-34$,
implying that by adopting such empirical baselines
$E_t=-34$ is effectively adopted. In that case, the two chain conformations
shown in Fig.~6 
would belong to the native state and therefore may be regarded as
sitting ``below the calorimetric baseline.''$^{19}$
These chains may model
the sparsely populated conformations revealed by native-state HX.$^{52-54}$
A multiple-conformation native state view 
is supported by a recent experimental observation of pretransitional 
conformational changes in ribonuclease A.$^{76}$

\section{\bf DISCUSSION}

\noindent
The logic of the present analysis is premised on a comparison between 
simulated folding/unfolding rates and transition-state 
predictions based on independently obtained free energy profiles.
The conventional transition state picture of folding posits a weak
or nonexistent dependence of the front factor on
a protein's intrachain interaction strength. In the present model, which
exhibits a chevron rollover, we find that the conventional picture holds 
approximately for unfolding but not for folding. In particular,
for the quasi-linear part of the folding arm of the chevron plot,
the folding front factor adopts a re-scaled form of the exponential
factor, harboring an exponent opposite in sign to that of the activation 
term. These findings are consistent with internal friction and diffusive 
folding dynamics ideas from energy landscape theory. They suggest that
simple transition state theory with a constant front factor may not
be generally applicable in the presence of a chevron rollover,
even if the kinetics is apparently first order. In this regard, future 
single-molecule measurements$^{77}$ of folding time distributions 
may provide important insight into the physics underlying approximate 
single-exponential folding kinetics, since these measurements may detect 
small deviations from first-order kinetics$^{11}$ (e.g., 
a possible small non-exponential tail in the distribution) that would 
otherwise be difficult to ascertain from traditional measurements of 
ensemble-averaged folding rates.

Chevron rollovers have been rationalized by dead-time discrete 
intermediates$^{29}$
and by movements of the transition-state peak 
on broad activation barriers.$^{31}$ We have not been able 
to detect these proposed features in our model free energy profiles. 
Instead, the present results offer an alternate rationalization in terms 
of diffusive dynamics and an interaction-dependent folding front factor.
It follows that, in general, analyses that focus exclusively on free energy 
profiles may be incomplete. Inasmuch as chevron rollovers are a manifestation 
of an interaction-dependent front factor, as suggested here, experimental 
observations of significant mutational effects on rollover behavior$^{29}$
imply that mutations can have a significant 
effect not only on the free energy profile itself, but also on front 
factors not afforded by such profiles.

The present account of salient features of chevron rollover and 
native-state HX in terms of an essentially continuous energy 
distribution (Fig.~3) is similar in spirit to the recent idea that these 
features may originate from a ``burst phase continuum.''$^{32,33}$ 
However, the burst phase continuum view is based on postulated
free energy profiles, not free energy profiles derived from models with 
explicit chain representations. Further effort will be required to elucidate 
the relationship between the burst phase continuum and
the present chain-based perspectives, as there are apparent differences 
between the two. For instance, the present study suggests that some of the 
states detected by native-state HX are on the native side of the conformational 
distribution (Figs.~3 and 6) rather than on the denatured side as envisioned 
by the burst phase continuum scenario. 

In summary,
we emphasize that while the current study proposes a new 
physical rationalization for chevron rollover, it does not by itself rule out 
other mechanisms.  Chevron rollovers in real proteins may arise from a 
combination of effects. Obviously, the generality of the present 
interaction-dependent front factor scenario should be further tested 
using model proteins with non-helical native topologies as well as using 
geometrically more realistic off-lattice continuum models.$^{26,78}$ 

As for the relationship between generic features of folding/unfolding 
kinetics and thermodynamics of small globular proteins, 
the qualitative agreement between Fig.~3 and typical chevron rollover plots
for real proteins supports the idea that proteinlike thermodynamics
necessarily lead to proteinlike folding/unfolding kinetics.
A case in point is the folding kinetics of a set of 20-letter model 
sequences reported by Gutin et al.$^{67}$ Our test calculations show that 
random 
sequences of this particular 20-letter alphabet with additive contact
energies are not calorimetrically 
two-state (data not shown). Although much useful insight has been 
gained from them$^{10,14,68}$ (see also ref.~16), recent calculations$^{19}$
indicate that even some designed sequences in this 20-letter model are 
thermodynamically less cooperative than the present model.$^{20}$
Apparently, the folding kinetics of these 20-letter model sequences are 
less proteinlike as a result, in that their folding 
rates {\it decrease} when native stability is increased from the 
transition midpoint (Figs.~2, 5, 7 and 8 in ref.~67). 
This is because the maximum folding rates and the onset of drastic rollover
in these 20-letter models occur around the thermodynamic transition midpoint,
rather than under strongly native conditions as in the cooperative model 
studied here (Fig.~3 of the present work). Thus, the chevron trend 
predicted by these 20-letter models under transition midpoint through 
moderately folding conditions is opposite to that observed 
experimentally,$^{49,51}$ because experiments
almost invariably show an increasing folding rate when native stability 
is increased from the transition midpoint.

The present model's kinetics is proteinlike but not two-state. In this 
respect, it is reassuring that the exercise here fares no worse than 
Nature's. This is because a protein's calorimetric two-state 
cooperativity,$^{79}$ such as that of hen 
lysozyme,$^{63,71}$ barnase,$^{29}$ 
and ribonuclease A,$^{62}$ is no 
guarantee for two-state kinetics.$^{4,5}$ 
However, the present exercise 
does suggest that additional or alternate interaction mechanisms have to 
be discovered to account for the strictly two-state behavior of many small 
single-domain proteins. In that regard, it would be interesting to 
investigate the connection between the strictly two-state proteins' 
apparently nonglassy kinetics$^{80}$
and the possibility that their front 
factors might be minimally sensitive to the variation in intrachain 
interaction strength. 

\noindent
{\bf Acknowledgments}

We thank Alan Davidson, Julie Forman-Kay, Yuji Goto, Bob Matthews, and 
Tetsuya Yomo for helpful discussions,
and David Baker, Martin Gruebele, Walid Houry, Kevin Plaxco, and 
Peter Wolynes for their critical
reading of an earlier draft of this paper and their very useful 
suggestions. This work 
was supported by Medical Research Council of Canada grant no. MT-15323 
and a Premier's Research Excellence Award (Ontario). H. S. C. is a 
Canada Research Chair in Biochemistry.

%\vfill\eject

%\par\vfill\eject

%\noindent
%{\large\bf References}

%\kern -1.5cm

\end{multicols}
\end{document}